# Mapping bibliographic metadata collections: the case of OpenCitations Meta and OpenAlex

Elia Rizzetto [1], Silvio Peroni [1]

[1] *Research Centre for Open Scholarly Metadata, Department of Classical Philology and Italian Studies, University of Bologna, Bologna, Italy*

**Abstract**

This study describes the methodology and analyses the results of the process of mapping entities between two large open bibliographic metadata collections, OpenCitations Meta and OpenAlex. The primary objective of this mapping is to integrate OpenAlex internal identifiers into the existing metadata of bibliographic resources in OpenCitations Meta, thereby interlinking and aligning these collections. Furthermore, analysing the output of the mapping provides a unique perspective on the consistency and accuracy of bibliographic metadata, offering a valuable tool for identifying potential inconsistencies in the processed data.

**Keywords**

Bibliographic collection, entity mapping, OpenCitations, OpenAlex

## 1. Introduction

Open bibliographic metadata collections play a pivotal role in enabling reproducible studies in the fields of bibliometrics, scientometrics and science of science and permit transparent procedures in the context of research assessment exercises, thus enabling the implementation of norms and guidelines that intend to reform the research assessment around the world, such as the Coalition for Advancing Research Assessment (CoARA[2]). As the volume and diversity of scholarly publications continue to expand, the need for comprehensive and interoperable bibliographic databases becomes increasingly pronounced.

This study delves into the process of mapping entities between two important open bibliographic metadata collections, OpenCitations Meta [1] and OpenAlex [2]. These mapping processes are a critical step towards enabling researchers, institutions, and platforms to access and utilise information seamlessly across diverse collections. In our work, the primary objective of this mapping is to integrate OpenAlex internal identifiers into the existing metadata of bibliographic resources (BRs) in OpenCitations Meta, thereby interlinking and aligning these collections. This paper presents the results of the mapping and provides details on the methodology adopted to accomplish this task. By shedding light on the complexities inherent in aligning bibliographic metadata collections, we aim to contribute valuable insights into the challenges and opportunities associated with such endeavours.

Furthermore, the study investigates the mapping process's implications to assess the quality of the involved datasets. Analysing the output of the mapping provides a unique perspective on the consistency and accuracy of bibliographic metadata, offering a valuable tool for identifying potential inconsistencies in the processed data. The importance of such considerations lies in their capacity to enhance data quality, fortify interoperability, and foster a more cohesive scholarly metadata landscape.

The rest of the paper is structured as follows. In Section "Material and methods", we introduce the processed data and the mapping methodology. Then, in Section "Results", we present the result of the

EMAIL: elia.rizzetto@studio.unibo.it (E. Rizzetto); silvio.peroni@unibo.it (S. Peroni)
ORCID: 0009-0003-7161-9310 (E. Rizzetto); 0000-0003-0530-4305 (S. Peroni)

[2] https://coara.eu/

mapping analysis. Section "Discussions" discusses some of the most relevant outcomes, highlighting the broader implications of mapping large bibliographic metadata collections for data integration, quality enhancement, and improved interoperability within the scholarly domain. Finally, in Section "Conclusions", we conclude the paper by sketching out some future works.

## 2. Material and methods

The following subsections analyse multi-mapped and non-mapped BRs in more detail.

### 2.1. Data

The two collections involved in the mapping process are OpenCitations Meta (OC Meta) and the OpenAlex catalogue (henceforth, just OpenAlex). In particular, only a subset of the entities in both collections has been considered for the mapping, namely – following OpenCitations nomenclature – bibliographic resources (BRs) [3][4], i.e. journal articles, conference papers, journals, books, book chapters, etc. The specific versions of the dataset used for the analysis described in the present study are version 5 of OC Meta [5] and a snapshot of the OpenAlex database released on October, 18th 2023[3].

OC Meta is the OpenCitations [6] database collecting metadata of scholarly bibliographic entities. The metadata exposed by OC Meta includes the basic metadata describing the BRs involved as citing or cited entities in the OpenCitations collection of bibliographic citations, i.e. OpenCitations Index [7]. In particular, OC Meta stores known persistent identifiers for each BR (DOI[4], PMID[5], PMCID[6], ISSN[7], and ISBN[8]), the title, type, publication date, page interval, the venue of publication, and the volume and issue numbers if the venue is a journal. In addition, OC Meta contains metadata regarding the main actors involved in the publication of each BR, i.e. the names of the authors, editors, and publishers, and their persistent identifiers (ORCID[9] and Crossref ID[10]) where available. All entities in OC Meta are persistently identified by the OpenCitations Meta Identifier (OMID), and their properties and relations are specified in compliance with the OpenCitations Data Model (OCDM) [3][4]. Notably, OC Meta also tracks the changes in its data and provides provenance information using Linked Open Data technologies. All OC Meta data is published under a CC0 license, is made accessible online via REST API[11] and SPARQL endpoint[12], and periodical dumps can be downloaded in tabular format (CSV files) and RDF (JSON-LD files)[13]. JSON-LD and CSV files are produced from a triplestore storing the whole OC Meta graph.

OpenAlex is a collection of scholarly metadata curated and published by OurResearch[14], and initiated in response to the discontinuation of the Microsoft Academic Graph (MAG) [8]. It features five types of entities providing rich metadata: Works (such as journal articles, books, and datasets), Sources (i.e. where works are contained, such as journals, conferences, and repositories), Authors, Institutions, and Concepts. Metadata include external persistent identifiers (PIDs): DOI, PMID, PMCID, and MAG ID for Work entities (journal articles, proceeding papers, etc.); ISSN, Wikidata ID[15], MAG ID and Fatcat ID[16] for Source entities (journals, books, etc.). Within OpenAlex, entities are identified with a persistent ID scheme, i.e. the OpenAlex ID. Data is published under CC0 license and accessible via a REST API, a web-based GUI, or as downloadable snapshots of the whole database (JSON-Lines files) [2].

[3] https://openalex.s3.amazonaws.com/RELEASE_NOTES.txt
[4] https://www.doi.org/the-identifier/what-is-a-doi/
[5] https://pubmed.ncbi.nlm.nih.gov/
[6] https://www.ncbi.nlm.nih.gov/pmc/about/public-access-info/
[7] https://www.issn.org/understanding-the-issn/what-is-an-issn/
[8] https://www.isbn-international.org/content/what-isbn/10
[9] https://info.orcid.org/what-is-orcid/
[10] https://www.crossref.org/
[11] https://w3id.org/oc/meta/api/v1
[12] https://w3id.org/oc/meta/sparql
[13] https://w3id.org/oc/download
[14] https://ourresearch.org/
[15] https://www.wikidata.org/wiki/Wikidata:Identifiers
[16] https://fatcat.wiki/

In the scope of this paper, the most relevant differences between OC Meta and OpenAlex concern the number of BRs in the two collections, the data sources they use, and some differences in the data models:

- OpenAlex is the largest open scholarly data collection, currently comprising 246,844,573 Works and 249,408 Sources, for a total of 247,093,981 BRs. The latest version of OpenCitations Meta includes 105,953,699 BRs.
- Data in OpenAlex is provided mainly by Crossref and inherited by the now-ceased Microsoft Academic Graph, but it also includes data from PubMed [9], the Directory of Open Access Journals (DOAJ) [10], Unpaywall [11], arXiv [12], Zenodo [13], the ISSN International Centre[17], and the Internet Archive's General Index[18]. OC Meta's sources are Crossref, the National Institute of Health Open Citation Collection (NIH-OCC, providing PubMed data) [14], OpenAIRE [15], and the Japan Link Center (JaLC) [16][19].
- In OpenAlex, Works can only have one ID value per each ID scheme, and Sources admit a list of up to two ISSNs or a single literal value for each of the other ID schemes. On the contrary, in the OCDM, and therefore in OC Meta, there are no limits on the number of possible values for each ID scheme. This substantial difference in how the two collections represent their data implies that, for example, if a journal article has been assigned two DOIs, they can be linked to the same entity (and the same OMID) in OC Meta, but not in OpenAlex. Another noteworthy difference is that OpenAlex does not support ISBNs, while OC Meta does.

## 2.2. Mapping process

The process leading to the mapping of these two collections is explained as follows. Initially, two tables are produced, which contain the internal IDs of the collections to be mapped with each other. The first table is produced by parsing the CSV dump of OC Meta, and, for each row, contains the OMID, external PIDs, and type for each BR in OC Meta that has external PIDs. The other table, produced from the JSON-Lines copy of the OpenAlex database, links each external PID in OpenAlex to the OpenAlex ID to which it is associated.

The table containing OpenAlex data is converted into a local SQL database. Then, the table containing OC Meta BRs to be mapped is iterated line by line, and each PID associated with each entity is looked up in the database containing PID-OpenAlex ID associations. The result consists of three additional tables:

1. A table storing OMID, OpenAlex ID, and type of the BRs, if exactly one OpenAlex ID per OMID has been found;
2. A table storing OMID, OpenAlex IDs, and type of the BRs, if multiple OpenAlex IDs per OMID have been found (multi-mapped BRs);
3. A table storing OMID and type of the BRs, if no OpenAlex ID has been found (non-mapped BRs).

The primary purpose of the mapping is to enable the addition of OpenAlex IDs to other available external persistent identifiers (PIDs) among the metadata of bibliographic resources already existing in OC Meta. However, the potential uses of the outcome of this process go beyond the ingestion of new metadata, proving it to be a useful tool for gaining a deeper understanding of the quality of the collections involved and for helping to identify any problems and inconsistencies therein. For this reason, the results of the mapping process regarding multi-mapped BRs and non-mapped BRs are analysed quantitatively and qualitatively according to the methodology described in the following subsections.

---

[17] https://www.issn.org/

[18] https://archive.org/details/GeneralIndex

[19] The data provided by JaLC is not included in the dump version processed for the mapping described by the present work (v5), but is included in the latest version (v6).

## 2.3. Multi-mapped BRs analysis: methodology

The mapping revealed that mapped entities in different datasets might go beyond a simple 1 to 1 alignment. Indeed, it is possible that one BR in OC Meta shares one or more external PIDs with more BRs in OpenAlex. These cases will be referred to as *multi-mapped* BRs.

Such cases, after being saved separately from the rest of the results, have first been checked manually by investigating sample resources, inspecting their full metadata in both datasets, making use of external APIs (Crossref [17] and DataCite [18]) and accessing the documents' location on the web via their PIDs. This study led to proposing an ad hoc categorisation, to frame the causes of such multi-mapping scenarios. We applied such categorisation to the instances of multi-mapped BRs by using heuristics to understand which category applies to the specific case.

The categories for OC Meta BRs that are multi-mapped to OpenAlex Works are the following:

1. *Category A* includes cases where two or more Works among the ones that are multi-mapped to a single OC Meta BR share at least one external PID. Given that external PIDs, such as DOIs, should be uniquely assigned to a BR, having more than one entity with the same external PID in the OpenAlex dataset means that there are either duplicate entities or errors in the metadata.
2. *Category B* includes cases where the same entity in OC Meta is mapped to different versions of the same publication, each represented by a Work entity in OpenAlex – e.g. in the case of having a version of record and one or more preprint and/or postprint versions. Preprints and postprints are hosted in a preprint server or a digital repository. DOIs of preprints or postprints are determined by considering the DOI prefix and looking it up on a list of DOI prefixes reserved for institutions that manage preprint servers or digital repositories for non-peer-reviewed publications.
3. *Category C* includes cases where the same entity in OC Meta is mapped to exactly 2 different Works in OpenAlex, and neither is a preprint or postprint version. The most likely causes for this scenario are errors in the data source used by OC Meta, bugs in OC Meta software, or different DOIs intentionally linked to the same OC Meta entity.
4. *Category D* includes cases where the same entity in OC Meta is mapped to multiple preprint versions of the same publication, each represented by a Work entity in OpenAlex. This typology is similar to category *B*, but it only includes preprint versions and detects them by checking for version number (e.g. "/v1") in the DOI value.
5. *Category E* includes cases where the same entity in OC Meta is mapped to multiple preprint versions of the same publication, each represented by a Work entity in OpenAlex. This typology is similar to categories *B* and *D*, but detects preprint versions by analysing the DOI value and checking if it contains semantic indicators that associate the DOI with a preprint server (e.g. "/arxiv" or "/zenodo").
6. *Category F* includes cases where the multi-mapped OpenAlex Works include a version of record, together with one or more Works of type "peer-review", "letter", "editorial", "erratum", or "other". For example, the DOI for an *erratum* notice and a DOI for the journal article that is being corrected may be wrongly assigned the same OMID in OC Meta, due to errors in the data source.

OC Meta BRs that are multi-mapped to OpenAlex Sources fit only into one category, "A", which groups cases where two or more multi-mapped OpenAlex Sources share at least one ISSN.

The categorisation process (represented as pseudocode in Listing 1) takes as input:

1. Multi-mapped BRs in the form of a table where each row represents the association of one BR in OC Meta with *n* BRs in OpenAlex, storing an OMID in the `omid` field and a list of OpenAlex IDs in the `openalex_id` field;
2. A list of 80 DOI prefixes that are assigned by Crossref and DataCite to organisations or institutions that manage preprint servers or digital repositories hosting non-peer-reviewed versions.
3. A list of strings that, when found inside a DOI value, indicate that the associated publication is hosted in a preprint server (e.g. "/arxiv", "/preprints", "/osf.io").
4. A SQL database storing full metadata of the OpenAlex BRs involved in the multi-mapping.

The process differentiates between OpenAlex Works and OpenAlex Sources. For rows storing Works, the process includes querying the database for external PIDs associated with each Work. If any PID is associated with multiple Works in the row, the categorisation is labelled with “A Subsequently, each multi-mapped Work is examined. If version-marked DOIs are present, the categorisation is labelled with “D. Otherwise, an assessment is made for DOI prefixes associated with preprint servers, leading to categorisations such as “B” for preprint server association, “E” for preprint indicators, “F” for meeting specific OpenAlex database criteria, and “C” for rows with only two Works.

For rows storing Sources, the process involves querying the database for ISSNs associated with each Source. If any ISSN is associated with multiple Sources in the row, the categorisation is labelled with “A”.

Rows that remain unclassified after these steps are marked as unclassified.

**Listing 1**

Pseudocode representing the process for multi-mapped categorization.

```
FUNCTION categorizationProcess(table, doiPrefixes, preprintIndicators, database):
  FOR EACH row IN table:
    IF Works IN row.openalex_id:
      externalPIDs = queryDatabaseForExternalPIDs(row)
      IF hasDuplicates(externalPIDs):
        row.category = "A"
      ELSE:
        FOR EACH work IN row.openalex_id:
          IF work.hasDOIs():
            IF hasVersionMarkedDOI(work, versionedDOIregex):
              row.category = "D"
            ELSE IF isPublishedByPreprintOrganization(work, doiPrefixes) AND
                (work.isSubmittedVersion() OR work.isAcceptedVersion()):
              row.category = "B"
            ELSE IF containsPreprintIndicator(work, preprintIndicators):
              row.category = "E"
            ELSE IF allDOIsHaveSamePrefix(work):
              IF work.isPeerReview() OR work.isEditorial() OR
                work.isErratum() OR work.isLetter():
                row.category = "F"
              ELSE IF countWorksInRow(row) == 2:
                row.category = "C"
    ELSE IF Sources IN row.openalex_id:
      issns = queryDatabaseForISSNs(row)
      IF hasDuplicates(issns):
        row.category = "A"
      ELSE:
        row.category = "non classified"
```

## 2.4. Non-mapped BRs provenance analysis: methodology

The results of the mapping process also include the resources that have not been mapped, since they also can provide useful insights on the nature of the processed data. In particular, non-mapped BRs are analysed with respect to their provenance information in OC Meta, specifically the primary source they have been derived from (Crossref, DataCite, NIH-OCC, JaLC, OpenAIRE). This analysis is performed by programmatically examining the RDF data including provenance information of all entities in the OC Meta collection, and considering only the nodes concerning non-mapped BRs. For each of these entities, we may have one or more primary sources depending on the number of times the entities' metadata have been modified, and on the source used as raw data provider for implementing such modifications For instance, if metadata information of a journal article was initially provided by Crossref during the first ingestion into OC Meta, and additional information was subsequently found for it from DataCite during a later data ingestion, both of these sources will be considered for the present analysis.

The provenance analysis process then counts the number of BRs for each source (or set of sources, in the case of resources originating from multiple sources) and for each type of BR (e.g. journal article, book, etc.). It was also decided to separate the counts based on the presence or absence of external PIDs to ensure additional granularity and significance of the results. Indeed, if an entity in OC Meta is not associated with any IDs other than OMID, it cannot be mapped to OpenAlex.

## 3. Results

Table 1 shows the number of processed BRs for both datasets and the general results of a quantitative analysis of the mapping output. As mentioned above, a BR entity in OC Meta can be mapped to a BR entity in OpenAlex only if both entities are associated with at least one external PID in common. Thus, the BRs in the OC Meta CSV dump that are theoretically mappable to at least one entity in OpenAlex are 90,270,131, and the set of OpenAlex BRs to which an OC Meta BR can be mapped amounts to 159,039,790 resources. Of the 90,270,131 mappable resources in the OC Meta CSV dump, most (approximately 97%) map to at least one resource in OpenAlex. However, a small number of these (173,513, roughly 0.2%) align (i.e. share external PIDs) with more than one entity in OpenAlex (multi-mapped BRs). At the same time, and vice versa, there is a consistent number of BRs *in OC Meta* (5,722,979) that do not uniquely map to a BR in OpenAlex, meaning that there are also cases where two or more BRs in OC Meta are aligned with the same entity in OpenAlex. These latter cases will be referred to as *inverted multi-mapped* BRs. Finally, 18,133,712 BRs in OC Meta do not map to any resource in OpenAlex, whether because, after being processed, they have been found not to have any corresponding entity in OpenAlex despite having external PIDs (2,963,534 BRs); because they do not have any external PID (9,000,386 BRs); or because they are not included in the CSV dump files, thus were not processed. Concerning the latter scenario, it is worth mentioning that the OC Meta software, when producing CSV dump files from the triplestore, does not represent journal issues and journal volumes as table rows. However, almost all BRs of these types lack external PIDs, with their OMID being the only persistent identifier.

**Table 1**

Number of processed, mapped, multi-mapped and non-mapped bibliographic resources.

| ***OC Meta*** | |
|---|---|
| Total No. of BRs in triplestore | 105,953,699 |
| No. of processed BRs (stored in CSV files) | 99,270,517 |
| No. of processed BRs with PIDs also supported by OpenAlex (stored in CSV files) | 90,270,131 |
| ***OpenAlex*** | |
| Number of BRs in dump | 245,207,435 |
| Number of BRs with PIDs supported also by OC Meta | 159,039,790 |
| ***Mapping OC Meta → OpenAlex*** | |
| No. of BRs in OC Meta mapped to exactly one BR in OpenAlex (1:1) | 87,605,238 |
| No. of BRs in OC Meta, which map to the same BR in OpenAlex as at least one other BR in OC Meta ($n$:1, *where* $n$>1) | 5,722,979 |
| No. of multi-mapped BRs in OC Meta (1:$n$, *where* $n$>1) | 173,513 |
| No. of non-mapped BRs in OC Meta | 18,133,712 |

### 3.1. Multi-mapped entities

Multi-mapped BRs have been analysed with respect to the number of OpenAlex entities mapped to a single BR in OC Meta. As shown in the distribution histogram in Figure 1, most cases involve two OpenAlex IDs per OMID (91.5%), followed by cases involving 3 OpenAlex IDs per OMID at a much lesser proportion (6.2%). The remaining cases (more than 3 OpenAlex IDs per OMID) are significantly less frequent, with values lower than 1.3%. It should also be mentioned, though, that some multi-mapped BRs are connected to a particularly high number of OpenAlex IDs: there are isolated cases of OC Meta BRs being mapped to more than 100 entities in OpenAlex, and even an outlier case involving 1,051 OpenAlex IDs. Such examples, though not common, may also help reveal potential anomalies or inconsistencies in both datasets.

Table 2 and Table 3 show the results of the categorisation of multi-mapped BRs, grouped by the type specified in OC Meta, involving OpenAlex Works and Sources, respectively. As concerns Works, most cases remain unclassified[20]. Nonetheless, we notice that BRs types that are most frequently involved in multi-mapping are journal articles, books, book chapters, resources whose type is not specified, and proceedings articles. Most cases, among the ones it was possible to classify, concern journal articles: publications for which the same PID (e.g. DOI) is assigned to multiple entities in OpenAlex (category A) and publications that are represented in 2 different Work entities in OpenAlex (category C). Other common cases for journal articles involve their publication in different versions: the preprint and/or postprint version, and possibly the version of record, are all merged into the same entity in OC Meta (categories B, D, and E). Other noteworthy cases involve book chapters in OC Meta mapped to 2 OpenAlex Works (category C) and resources of unspecified type assigned version-marked DOIs (category D).

Regarding Sources, the most common case involves journals for which the same ISSN is attributed to more than one entity in OpenAlex (category A). Non-classified Sources, mostly journals, are likely caused by OpenAlex not associating different ISSNs to the same journal entity. Journals, indeed, can be assigned two different ISSNs, one for the print version and one for the online version; sometimes they can even receive more than two ISSNs, if for example there have been changes in the journal name. While OC Meta tends to prioritise the fundamental continuity of the journal entity – regardless of variations in names, the number of ISSNs, or diverse publication media – OpenAlex occasionally encounters challenges in consolidating all ISSNs under a single entity. In Example 1, for the journal identified as "br/06602375171", the "Journal of Health"[21], OC Meta has two ISSNs, each assigned to a different entity in OpenAlex (S2764583335, associated with the online ISSN, and S4210187171, associated with the print ISSN).

| omid | openalex_id |
| --- | --- |
| br/06602375171 | S2764583335 S4210187171 |

(Example 1)

**Table 2**
Number of multi-mapped OC Meta BRs for each BR type and category. Cases involving OpenAlex Work entities.

| *OC Meta br type* | *A* | *B* | *C* | *D* | *E* | *F* | *Unclassified* |
| --- | --- | --- | --- | --- | --- | --- | --- |
| **Total: 167054** | **39,758** | **9,421** | **3,8984** | **12,496** | **1,376** | **887** | **64,132** |
| journal article | 38,179 | 8,722 | 35,744 | 10,196 | 1,030 | 805 | 50,579 |
| book | 27 | 1 | 581 | 31 | 0 | 4 | 8,511 |
| book chapter | 341 | 8 | 1,112 | 21 | 4 | 36 | 2,002 |
| <unspecified> | 607 | 502 | 609 | 1,753 | 265 | 29 | 1,503 |

[20] These could potentially include mappings that the categorization heuristics failed to catch, or concern general errors in the data sources used by OC Meta and/or OpenAlex.

[21] https://journal.gunabangsa.ac.id/index.php/joh/

| | | | | | | | |
|---|---|---|---|---|---|---|---|
| proceedings article | 477 | 10 | 452 | 108 | 16 | 0 | 666 |
| proceedings | 8 | 24 | 230 | 13 | 1 | 0 | 508 |
| report | 13 | 1 | 155 | 1 | 0 | 0 | 167 |
| reference book | 0 | 0 | 7 | 0 | 0 | 0 | 69 |
| reference entry | 99 | 7 | 22 | 0 | 1 | 13 | 57 |
| web content | 2 | 146 | 14 | 335 | 58 | 0 | 47 |
| dataset | 1 | 0 | 38 | 37 | 0 | 0 | 10 |
| dissertation | 0 | 0 | 9 | 1 | 1 | 0 | 9 |
| series | 0 | 0 | 4 | 0 | 0 | 0 | 3 |
| standard | 0 | 0 | 6 | 0 | 0 | 0 | 1 |
| book section | 0 | 0 | 1 | 0 | 0 | 0 | 0 |
| journal | 4 | 0 | 0 | 0 | 0 | 0 | 0 |

**Table 3**

Number of multi-mapped OC Meta BRs for each BR type and category. Cases involving OpenAlex Source entities.

| *OC Meta br type* | *A* | *Unclassified* |
|---|---|---|
| **Total: 6459** | ***4,076*** | ***2,383*** |
| journal | 4,057 | 2,345 |
| book series | 17 | 38 |
| series | 2 | 0 |

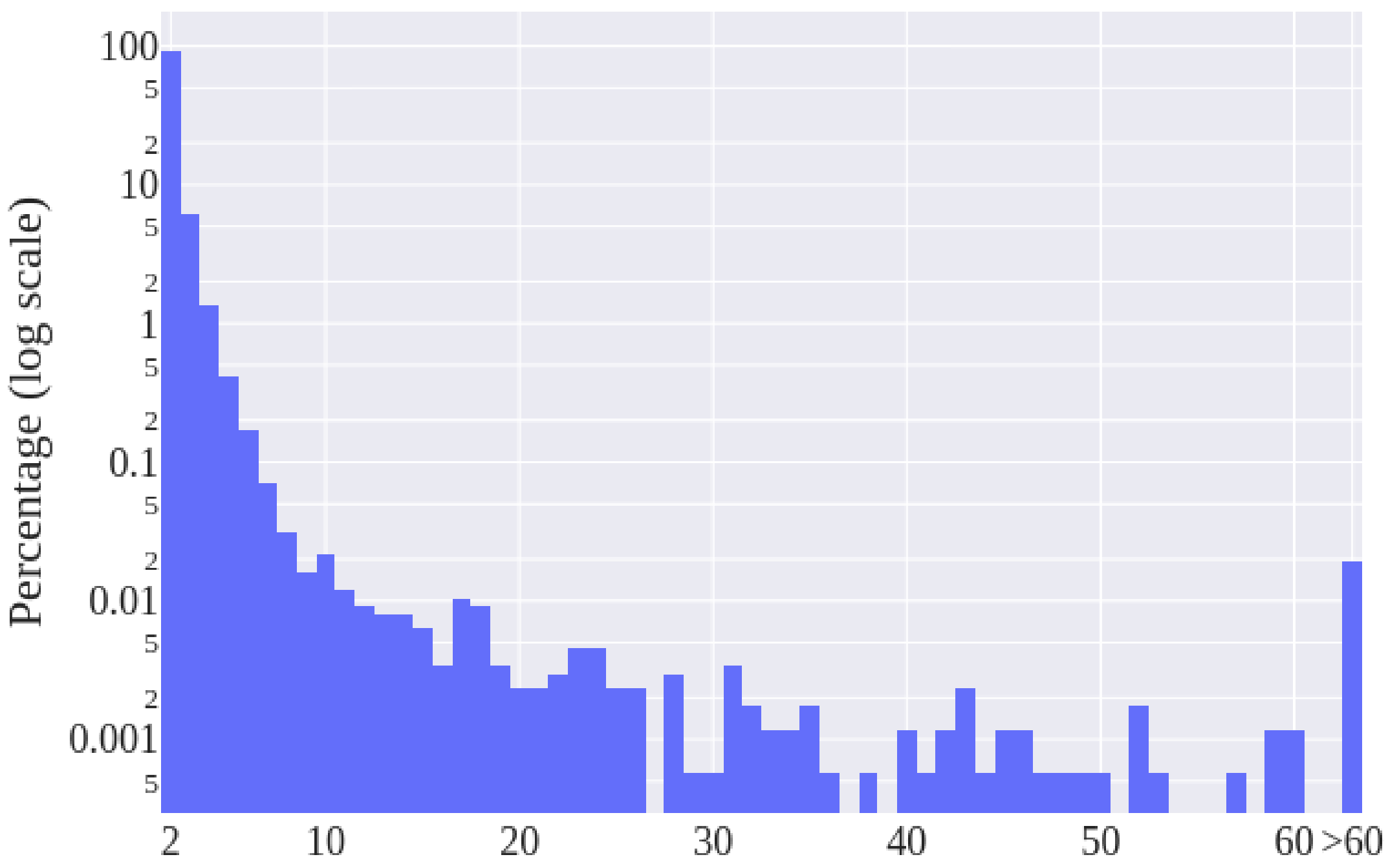

**Figure 1**: Histogram representing the distribution of multi-mapped OMIDs by the number of the OpenAlex IDs found for a single OMID.

## 3.2. Non-mapped entities

The entities in OC Meta that have not been mapped to any entity in OpenAlex (i.e. non-mapped entities) have been analysed with regard to the source they have been provided by. Provenance information is available as RDF data for the great majority of non-mapped entities, with only 2094 being left out. Approximately 83% of non-mapped entities do not have any other PID than their OMID, therefore they cannot be mapped until any other PID also supported by OpenAlex is associated with them in OC Meta data. Table 4 illustrates a representative sample of the results of provenance analysis, concerning the ten most frequent bibliographic entity types among non-mapped entities: it shows how many non-mapped entities derive from each source or set of sources, and entities are grouped by the type of BR and by the presence/absence of other PIDs besides OMID.

**Table 4**
Number of non-mapped OC Meta BRs for each provenance source and BR type. The column "External PID?" indicates whether the values in the row refer to BRs for which other PIDs than OMID are registered in OC Meta. The intersection symbol (∩) connecting two data sources indicates that the counts on the row refer to BRs for which the OC Meta provenance data provides multiple data sources across the snapshots.

| | *External PIDs?* | *proceedings* | *journal issue* | *book* | *journal volume* | *dataset* | *unspecified* | *journal article* | *reference book* | *report* | *journal* |
|---|---|---|---|---|---|---|---|---|---|---|---|
| **total by** | - | 5,383,11 | 5,064,03 | 2,521,88 | 1,576,74 | 1,242,10 | 1,419,21 | 253,284 | 188,453 | 135,997 | 103,263 |

| **type →** | | 5 | 0 | 6 | 4 | 1 | 2 | | | | |
|---|---|---|---|---|---|---|---|---|---|---|---|
| *Crossref* | *no* | 5,370,793 | 4,870,258 | 2,407,893 | 1,547,907 | 0 | 428,182 | 0 | 188,426 | 0 | 61,561 |
| | *yes* | 31 | 79,934 | 108,658 | 95 | 46 | 473,428 | 5,467 | 26 | 15 | 55 |
| *Zenodo* | *no* | 0 | 1,075 | 0 | 0 | 0 | 0 | 0 | 0 | 0 | 0 |
| | *yes* | 11,487 | 0 | 1,247 | 0 | 0 | 355,075 | 202,018 | 0 | 1,993 | 19 |
| *NIH* | *no* | 786 | 102,115 | 3,757 | 22,602 | 0 | 1 | 0 | 1 | 0 | 40,080 |
| | *yes* | 0 | 0 | 0 | 0 | 0 | 153 | 42,009 | 0 | 0 | 1,499 |
| *Datacite* | *no* | 0 | 2 | 0 | 1 | 0 | 9 | 0 | 0 | 0 | 1 |
| | *yes* | 0 | 0 | 300 | 0 | 1,238,173 | 162,061 | 1075 | 0 | 133,865 | 6 |
| *Zenodo ∩ Crossref* | *no* | 0 | 2,830 | 0 | 1,730 | 0 | 0 | 0 | 0 | 0 | 5 |
| | *yes* | 17 | 37 | 2 | 0 | 190 | 16 | 57 | 0 | 0 | 0 |
| *Datacite ∩ Crossref* | *no* | 1 | 457 | 1 | 436 | 0 | 0 | 0 | 0 | 0 | 0 |
| | *yes* | 0 | 23 | 28 | 0 | 3,521 | 287 | 7 | 0 | 115 | 0 |
| *NIH ∩ Zenodo* | *no* | 0 | 3,847 | 0 | 910 | 0 | 0 | 0 | 0 | 0 | 13 |
| | *yes* | 0 | 0 | 0 | 910 | 0 | 0 | 374 | 0 | 0 | 21 |
| *NIH ∩ Crossref* | *no* | 0 | 3,307 | 0 | 2,125 | 0 | 0 | 0 | 0 | 0 | 0 |
| | *yes* | 0 | 120 | 0 | 0 | 0 | 0 | 2,246 | 0 | 0 | 3 |
| *NIH ∩ Zenodo ∩ Crossref* | *no* | 0 | 12 | 0 | 27 | 0 | 0 | 0 | 0 | 0 | 0 |
| | *yes* | 0 | 13 | 0 | 0 | 0 | 0 | 31 | 0 | 0 | 0 |
| *Zenodo ∩ Datacite ∩ Crossref* | *no* | 0 | 0 | 0 | 1 | 0 | 0 | 0 | 0 | 0 | 0 |
| | *yes* | 0 | 0 | 0 | 0 | 6 | 0 | 0 | 0 | 3 | 0 |
| *Zenodo ∩ Datacite* | *no* | 0 | 0 | 0 | 0 | 0 | 0 | 0 | 0 | 0 | 0 |
| | *yes* | 0 | 0 | 0 | 0 | 165 | 0 | 0 | 0 | 6 | 0 |

## 4. Discussion

The mapping process and the analysis of its results concerned the study and use of a great amount of data from the involved databases, requiring, for example, the consideration of all bibliographic entities in their entirety. This study highlighted problems and inconsistencies within the used datasets. First, concerning OC Meta, the process provided an opportunity to conduct counts of the number of entities contained in the CSV and JSON-LD files comprising the dump. This highlighted a discrepancy between the number of BRs contained in the triplestore and the number of BRs actually reported in the dump files constructed from the triplestore. Additionally, it was observed that this numerical difference is also reflected in the RDF files containing provenance information.

The analysis of multi-mapped and the count of inverted multi-mapped BRs posed interesting questions as well. A comparison between OC Meta and OpenAlex from the perspective of two different data models helped emphasise that both collections have duplicate entities, i.e. resources sharing the same external PID (e.g., DOI or ISSN, which should be uniquely assigned), with at least one other resource within the collection. In the case of multi-mapped BRs, it was further found that the alignments of a single OMID to multiple OpenAlex IDs could be attributed partly to natural diversities between data models, partly to errors in data sources, and partly to errors in the software used to populate the collection. Generally, OC Meta tends to erroneously group various expressions of a resource (preprints, postprints, and versions of record) into a single entity, propagating errors present in data sources, even when there should be two separate entities (e.g. in the case of a version of record and its preprint). In contrast, OpenAlex generally tends to have separate entities due to limits on the number of possible values for each ID scheme and more intensive data correction activities made possible by the use of web crawlers.

From the perspective of OC Meta, while some of these multi-mapped cases result from data representation choices, others are the result of errors often originated from sources (especially in cases where an OMID is aligned to a very high number of OpenAlex IDs).

Regarding non-mapped BRs, we observed that, despite OpenAlex formally including a greater number of entities than OC Meta, approximately 5 million OMIDs are not associated with any corresponding OpenAlex ID. This is partly because some of the resources counted as non-mapped BRs (15,170,179 BRs) were not included in the CSV files that the mapping process takes as input; therefore they were not processed at all during the mapping phase. Of the other 2,963,533 non-mapped resources, those with one or more external PIDs are particularly interesting, as one would expect them to have at least one corresponding entity in OpenAlex.

In this regard, it should be noted that, in the case of the 108,658 non-mapped books from Crossref, many resources likely have only ISBNs among the external IDs, which are not supported by OpenAlex and therefore cannot be used for mapping. Another interesting case is the set of dataset resources from DataCite, totalling 1,238,173 entities, which can be explained by the fact that DataCite is not among the sources used by OpenAlex. More generally, the 15,061,152 non-mapped BRs without external PIDs underscore the unique contribution made by OC Meta by assigning a persistent identifier, i.e. OMID, to entities that would otherwise lack one. Indeed, the OCDM permits to represent journal issues and journal volumes as first-class entities, while they are typically represented only as metadata associated with journal articles (as is the case for OpenAlex).

## 5. Conclusions

The results of the mapping of OpenCitations Meta bibliographic resources to OpenAlex bibliographic resources have provided valuable insights into the integration of bibliographic metadata entities, showcasing that the majority of processed OC Meta resources are successfully mapped with exactly one entity in OpenAlex. This achievement is significant, as it allows for the direct ingestion of OpenAlex IDs into the metadata of the corresponding bibliographic resources in OC Meta. This seamless integration enhances the interconnectedness and interoperability of these two substantial bibliographic collections.

However, challenges were encountered in the case of multi-mapped BRs, leading to the decision to temporarily exclude them from being included in OC Meta. While this choice poses a limitation, the analysis of these multi-mapped entities has proven instrumental in identifying inconsistencies within both datasets. Furthermore, the examination of non-mapped resources, considering their type and provenance, has underlined the impact of using different data sources and different identifiers in the collections to map, resulting in quite a significant limitation of the mapping coverage.

Addressing the limits and inconsistencies revealed by the mapping results, OpenCitations has proactively taken measures to rectify errors and enhance the quality of its data, particularly in the production process of dump files. Future developments are envisioned, e.g. to further refine the management of scenarios involving bibliographic resources being associated with multiple values for the same ID scheme (e.g. multiple DOIs for the same journal article). Improvements like these aim to bolster the robustness of the mapping process as well as the quality of the data, ensuring a more accurate and comprehensive representation of bibliographic entities.

## 6. Acknowledgements

This project has been partially funded by the European Research Council Executive Agency under service contract ERCEA/2023/VLVP/0007 and the European Union's Horizon Europe framework programme under grant agreement No. 101095129 (GraspOS Project).